\documentstyle[aps,epsfig]{revtex}
\epsfverbosetrue
\begin{document}
\draft

%\documentstyle[twocolumn,aps,epsf]{revtex}
%\documentstyle[prc,aps]{revtex}
%\begin{document}
%\draft

\title{Soft Physics, Centrality and Multiplicity at RHIC}  

\author{D.~E.~Kahana, S.~H.~Kahana}
 
\address{Physics Department, Brookhaven National Laboratory\\
   Upton, NY 11973, USA\\}
%    $^{2}$31 Pembrook Dr., \\
%   Stony Brook, NY 11790, USA}
\date{\today}  
  
\maketitle  
  
\begin{abstract}

The  inclusive spectra  so  far accumulated  at the  Relativistic
Heavy Ion Collider (RHIC) at energies of $\sqrt{s} = 56, 130$ and
$200$ GeV  are examined  within the hadronic  simulation LUCIFER.
What  emerges  at this  juncture  is  a  comprehensive and  clear
picture   of  soft   physics  which   apparently   dominates  the
intermediate and later stages of the ion collisions. The focus is
on energy  and centrality dependence of  the mid-rapidity charged
spectra, using an analysis based  for the most part on production
and rescattering of intermediate generic resonances. The bosonic,
$\rho$-   and  $K^{*}$-like   resonances,  produced   in  initial
nucleon-nucleon  interactions and  materialising only  after some
delay  time, behave  as  an incompressible  fluid with  saturated
number and energy density.

\end{abstract}

\pacs{25.75, 24.10.Lx, 25.70.Pq}

\section{Introduction}  

A possible  conclusion to be  drawn from initial  observations of
inclusive charged particle spectra in Au + Au collisions at RHIC,
by   the    four   detectors   PHOBOS    \cite{phobos1},   PHENIX
\cite{phenix1},  BRAHMS  \cite{brahms1}  and  STAR  \cite{star1},
which observations, of course, are only of the final state of the
collision,  is  that  soft,  strongly  interacting  QCD  plays  a
dominant  role in  the  dynamics.   To be  sure,  there do  exist
potentially  interesting  anomalies  in  high  $p_\perp$  charged
\cite{star2}  and  neutral  \cite{phenix2}  pion  spectra,  among
others.  These  occur for $p_\perp$ in the  range $2$--$4$ GeV/c,
involve only  a small  fraction of the  integrated cross-section,
and may  in fact be describable  by an energy  loss mechanism for
$\pi$ mesons acting in the later stages of a collision.

The hadronic  simulation LUCIFER \cite{lucifer,lucifer1,lucifer2}
has  already  given a  good  description  \cite{lucifer2} of  the
earliest PHOBOS \cite{phobos1} and STAR \cite{star1} measurements
of  averaged mid-rapidity  cross-sections for  charged particles,
$(dN/d\eta)^{ch}$   for  $|\eta|  \leq   1$,  including   also  a
successful prediction \cite{lucifer1,phobos2} of the surprisingly
slight ($10$--$14$\%)  rise in  this observable as  the collision
energy  is increased from  $\sqrt{s}=130$ GeV  to 200  GeV.  Also
correctly  anticipated  was  the  shape  of  the  pseudo-rapidity
spectrum  for  charged  particles  at  higher  values  of  $\eta$
\cite{lucifer2,phobos2,brahms2}.

Here,   we  consider  simulation   results  for   the  centrality
dependence   which   is   found   by   all   of   the   detectors
\cite{star1,phenix2,brahms2,phobos3}.     These    collaborations
emphasize the  variation of $(dN/d\eta)^{ch}|_{\eta=0}$  with the
number   of  participants.   This   work  concerns   itself  with
comparisons of  the spectral magnitudes and shapes  as a function
of centrality. The variation of experimentally defined centrality
is  directly  related  theoretically  to changes  in  the  impact
parameter $b$.   The energy and  geometrical dependence extracted
from LUCIFER speak directly to assumptions underlying the cascade
simulation,  and to  the nature  and  the number  density of  the
resonance rescatterers employed in the model.

In earlier work \cite{lucifer,lucifer1,lucifer2} we stressed that
the role  of our  simulation was only  to define  the `ordinary,'
giving experimentalists a background calculation against which to
judge  potentially unusual  data. There  do, of  course,  exist a
variety              of             other             simulations
\cite{wang1,frithjof,ranft,werner,parton,geiger2,RQMD,URQMD,KO},
having similar  goals but  differing approaches: some  are purely
partonic models and some are hybrid partonic-hadronic.

It  seems again  appropriate to  compare the,  by  now, extensive
observations at  RHIC with simulations  which do not  assume that
plasma is necessarily present.   The purest such comparison would
employ a model involving only colour singlet, hadronic degrees of
freedom.  The physical picture is one in which ion-ion collisions
are  describable in  the  main by  multiple interactions  between
excited  hadrons only.   In such  a picture,  constituent quarks,
present within original and  produced baryons, are perhaps raised
to  states  differing  from  those  present in  the  lowest  mass
baryons,  but glue holding  the valence  quarks in  place remains
`sticky.'   So quarks  and gluons  continue  to act  as if  still
confined within  hadrons.  In addition to  excited baryons, color
singlet  bosonic  degrees of  freedom  are  produced  and play  a
significant role.  Such a  description has proven feasible in the
Pb + Pb collisions  examined in NA49 \cite{NA49,NA49b}.  In fact,
excited bosons  only enter  at a later  stage of  the simulation,
often   after  some   initial   baryon-baryon  interactions   are
completed.  It remains to be  seen whether at the higher energies
of RHIC a  large number of partons, carrying  a large fraction of
the energy density, perhaps `minijets' \cite{eskola}, are in fact
free  and able  to roam  over large  spatial distances,  and more
importantly  whether  sufficient  `free'  gluons are  present  to
create  the  thermodynamic  basis  for material  which  could  be
described as quark-gluon plasma.

The simulation we employ has  two successive stages.  In phase I,
no  energy  is removed  for  soft  processes. Therefore,  initial
baryon-baryon interactions  take place  very nearly on  the light
cone. The construction of a collision history is guided by inputs
from   NN  data,  the   key  ones   being:  (1)   absolute  total
cross-sections as a function  of energy, (2) branching ratios for
elastic  and  various inelastic  processes  and (3)  multiplicity
distributions for meson production.

Our two body model has been discussed extensively in earlier work
\cite{lucifer1,lucifer2}.      It     attempts     to     emulate
multi-peripheral  models   of  nucleon-nucleon  scattering  which
divide the  total cross-section into  elastic, single diffractive
and non-single diffractive \cite{multiperipheral} processes.  Our
assumption   is  that   generic   $\rho$-like  and   $K^{*}$-like
resonances are  the major  rescatterers, {\it only  entering into
phase II}, of the simulation, which begins after a formation time
$\tau_f$, a  central model  parameter, has elapsed  for particles
produced in phase  I, in their various rest  frames.  Since these
resonances themselves decay  into multiple particles stable under
the  strong interaction  on a  time  scale $\tau_d$,  which is  a
second  important model parameter,  their multiplicities  must be
adjusted appropriately so as to reproduce free space NN data, and
they are  in fact reduced  in number by  some two to  three times
relative   to  the  expected   multiplicity  of   stable  mesons.
Importantly however,  it then follows that  any model reproducing
the known  nucleon-nucleon data,  such as explicit  string models
\cite{frithjof,werner},  could  have  been  employed  up  to  the
commencement of phase II of the simulation.

We  use  a  resonance  model of  nucleon-nucleon  scattering  for
reasons  which  will  become  clear  as we  proceed  and  because
interactions which are particle-like in nature are easier to deal
with than, say, the interactions of strings.  We will discuss the
spatial and energy densities of the `rescatterers' at some length
in what  follows and thus make  evident a simplicity  in both the
energy and centrality  dependence of the observed pseudo-rapidity
densities which arises from such considerations.  It appears that
over an  appreciable range of  both energy and  impact parameter,
the density of generic hadronic resonances, built up during phase
I,  must remain  nearly constant  in ion-ion  collisions,  if the
observed  centrality  and  energy  dependence  of  final  charged
particle spectra  are to  be correctly described.   The extracted
resonance  size,   assuming  close  packing   or  saturation,  is
eminently reasonable.
 
We again adopt the position that the analysis for Au + Au at RHIC
simply  presents  an extrapolation  from  earlier NA49  inclusive
measurements   \cite{lucifer2,NA49,NA49b}  to   the  considerably
higher  energy  RHIC  collisions.   We could  equally  well  have
normalized the  absolute level of meson production  to the lowest
RHIC  energy ($\sqrt{s}=56$  GeV).   This normalization  directly
implies  a  determination  of  the  average  density  of  generic
resonances,  or   rescatterers,  in  phase  II   of  the  LUCIFER
simulation for Au + Au,  a theoretical observable which is then a
single  number characterising  presently measured  inclusive RHIC
physics.

Uniformity  in resonance  production is  achieved by  imposing an
upper  limit  on  total  multiplicity  in  contiguous  groups  of
interacting    nucleons     within    an    A     +    A    event
\cite{lucifer1,lucifer2}  or, equivalently,  by imposing  a limit
directly  on  the  two  body interactions  themselves.   This  is
accomplished by  introducing this limit  at a single  energy. The
density of the saturated  resonance fluid is fixed by normalising
the simulated multiplicity to  the experimental multiplicity at a
fixed total energy in a  massive (large A) ion-ion collision.  No
such  restriction is  imposed  on pure  NN  interactions, or  for
example p + D collisions.  Previously, \cite{lucifer,lucifer1} we
based   our  approach   on   ideas  put   forward  by   Gottfried
\cite{gottfried}  in  describing  p  + A  collisions.   Gottfried
delayed interactions  of particles produced in  the successive NN
interactions in a  p + A collision, until  the produced particles
were physically sufficiently separated  from each other and their
progenitors  to act  as separate  hadrons.  The  extra constraint
which we impose, saturation of the rescatterer density in ion-ion
collisions, may  be occasioned when  multiple resonances produced
in neighbouring NN collisions  interfere with one-another.  It is
almost  self evident that  individual resonances  condensing from
such two body collisions cannot overlap.

After a  short digression on  the details of LUCIFER,  we develop
the above ideas in  two ensuing sections.  First, comparisons are
presented between the model  and the experimental results at RHIC
from  the  various detectors,  large  and  small.  These  include
charged  particle pseudo-rapidity  measurements  at $\eta=0$  for
varying centrality,  at the first RHIC  energy $\sqrt{s}=130$ GeV
analysed in  this fashion. Also treated is  the energy dependence
for  $(dN/d\eta)^{ch}|_{\eta=0}$  at  $\sqrt{s}=56$,  $130$,  and
$200$ GeV, using for the  lowest energy, data which are available
only   from  the   PHOBOS  \cite{phobos1}   collaboration.   Also
considered  at $\sqrt{s}=130$ GeV,  are the  full pseudo-rapidity
distributions for differing degrees of centrality.

Secondly,  the  distributions  of underlying  rescatterers,  {\it
i.~e.~} the  dependence of  the number of  boson-like resonances,
$N_{RES}(b,s)$,  on impact  parameter and  energy,  are extracted
from the  simulations. The simple  relations for the  spatial and
energy distributions thus  obtained illuminate the physical basis
for the overall simulation.  In particular, the number density of
meson resonances reached at the formation time $\tau_f$, although
not  large, is sufficient  to explain  the observed  stable meson
production. The baryon  density from the onset of  an AA event up
to  the earliest  formation time  t=$\tau_f$, involves  far fewer
particles, but  could still  in principle provide  problems.  The
latter will be discussed at an appropriate point.

\section{LUCIFER: Dynamics and Inputs} 

For completeness  we present a  brief outline of the  dynamics in
LUCIFER  \cite{lucifer1,lucifer2}. Phase  I  has been  adequately
described above  and sets the initial conditions  for an ordinary
hadronic    cascade   (with    energy   loss)    in    phase   II
\cite{ARC1,RQMD,KO}.  The numbers  and  the properties  (flavour,
mass,  {\it  etc.}~)  of  hadrons  entering  into  phase  II  are
determined   in   phase   I   by  recording   all   baryon-baryon
collisions.  Providing complete initial  conditions for  phase II
requires  us  to  specify  four  momenta and  positions  for  the
produced   resonances.   The   positions  are   simply   randomly
distributed  along the paths  of nucleons  colliding in  phase I,
while the  momenta must conserve overall energy  and momentum and
be   consistent   with  production   and   phase   space  in   NN
collisions.  Existing  baryons  acquire  transverse  momentum  by
making  a random  walk in  the number  of collisions,  and softly
produced   meson   resonances   are  given   transverse   momenta
commensurate with  soft production as observed  in NN collisions.
The  total  multiplicity  of  the  resonances  is  restricted  as
described above. We reemphasise that the generic resonances first
appear only in phase II. The  latter is a normal cascade in which
energy  is lost  at each  binary collision  and during  which the
resonances both interact with  each other and decay into pions and
other stable  particles. The restricted  range of masses  for the
resonances, 300-1200  MeV, or  somewhat higher if  strangeness is
involved, guarantees  we have  only soft processes,  {\it i.~e.~}
that only relatively low $p_\perp$ is generated in the decays.

We have listed  the important two-body inputs in  phase I, but we
should  make  more  explicit  a  very important  feature  in  the
elementary  model,  {\it  viz}   the  energy  dependence  of  the
multiplicity  distributions, which  obey  KNO-scaling \cite{KNO}.
Our description of the  scaling of the multiplicity distributions
comes from  fitting multi-prong data mostly drawn  from the UA(5)
experiment \cite{UA(5)}.  Central to this parameterization is the
energy  dependent average  number of  prongs $\overline  {n} (s)$
entering through the definition of KNO scaling

\begin{equation}
n=\overline{n} (s) f(\frac{n}{\overline{n}(s)}).
\end{equation}

\begin{figure}
\vbox{\hbox to\hsize{\hfil
\epsfxsize=6.1truein\epsffile[0 0 561 751]{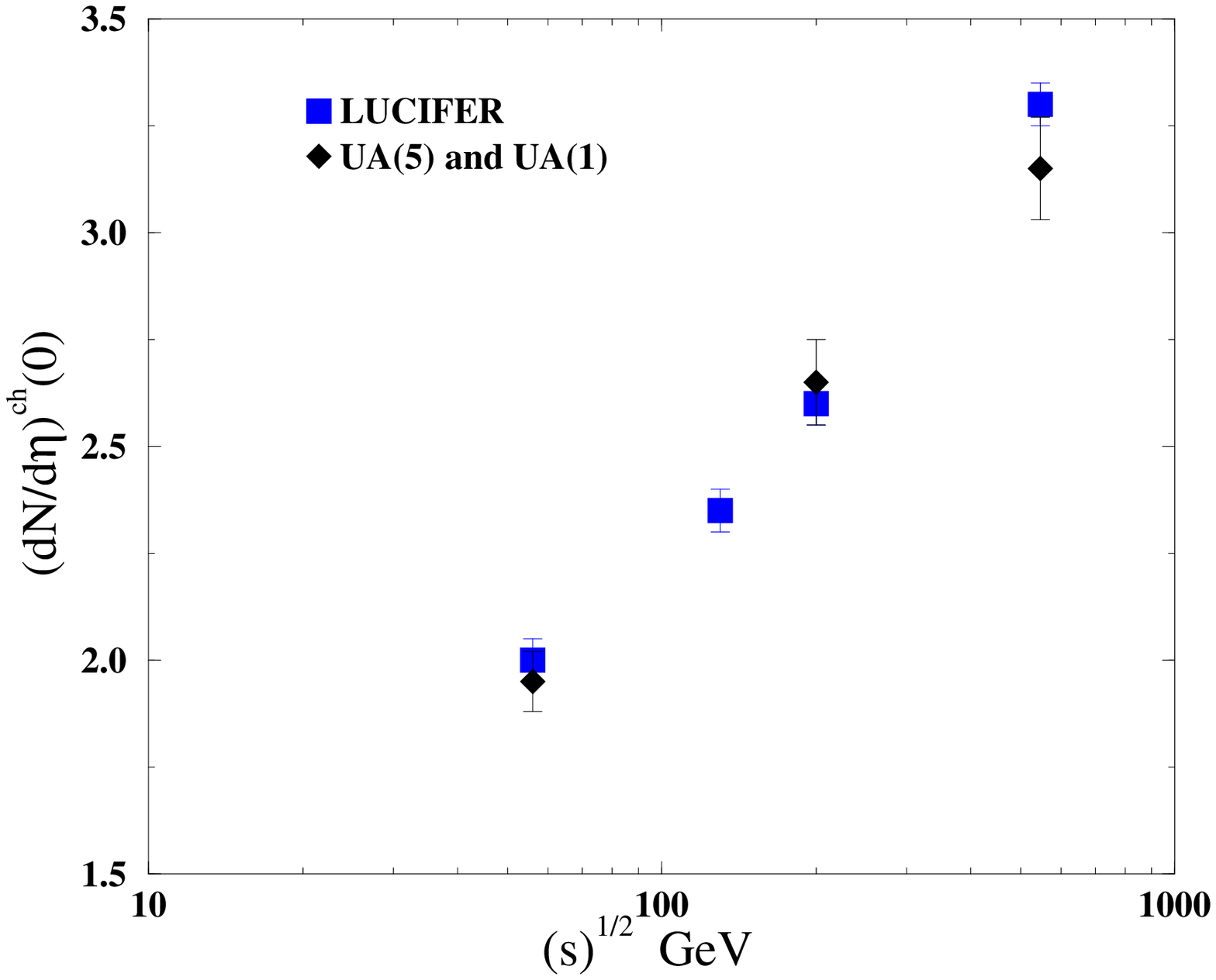}
\hfil}}
\caption[]{Proton/Anti-proton   charged   meson   densities   for
$\eta=0$ at energies relevant to RHIC vs those extracted in $p+D$
from  LUCIFER.  The agreement  with elementary  measurements from
UA(5) and UA(1) displayed is in fact a prediction of the two body
model  used   in  LUCIFER,  since   the  basic  inputs   are  the
multiplicity distributions from single and non-single diffractive
data and hence also the average multiplicity $\bar n(s)$ in
the KNO distributions assumed.}
\label{fig:ua5dndeta}
\end{figure}

Although we  artificially limit the number of  mesons produced in
ion-ion collisions  by our  no-overlap constraint on  the generic
resonances,  the energy dependence  for increasing  production of
these  particles  still  drives  the final  multiplicities  as  a
function  of  $s$. In  Figure  \ref{fig:ua5dndeta}  we display  a
De facto  prediction of  the two  body model,  {\it  i.~e.}~which
arises via  the energy dependence of  $\overline{n}(s)$.  This is
the comparison of calculated  energy variation of central charged
multiplicity,   $(dN/d\eta)^{ch}(s)|_{\eta   =   0}$  with   data
\cite{UA(5)2,CDF,UA(1)}.   Since  one  is  here  dealing  with  a
collision  involving two  incoming  nucleons (participants)  this
quantity is, in  effect, already divided by 1/2  times the number
of participants.   The slight increase predicted for  the rise in
$(dN/d\eta)^{ch}|_{\eta=0}$   for    Au   +   Au,    going   from
$\sqrt{s}=130$  GeV to $200$  GeV, arises  essentially completely
from the energy dependence  of $\overline{n}$: this dependence is
weak  in  the  two  body   data  and  equally  weak  for  ion-ion
collisions. This is a  dynamic, and not merely kinematic, outcome
of  the simulation,  but  it is  one  which follows  from the  NN
character of phase I.

\section{Centrality Dependence of Rapidity Distributions: Lucifer 
vs Data. } 

\begin{figure}
\vbox{\hbox to\hsize{\hfil
\epsffile[0 0 499 599]{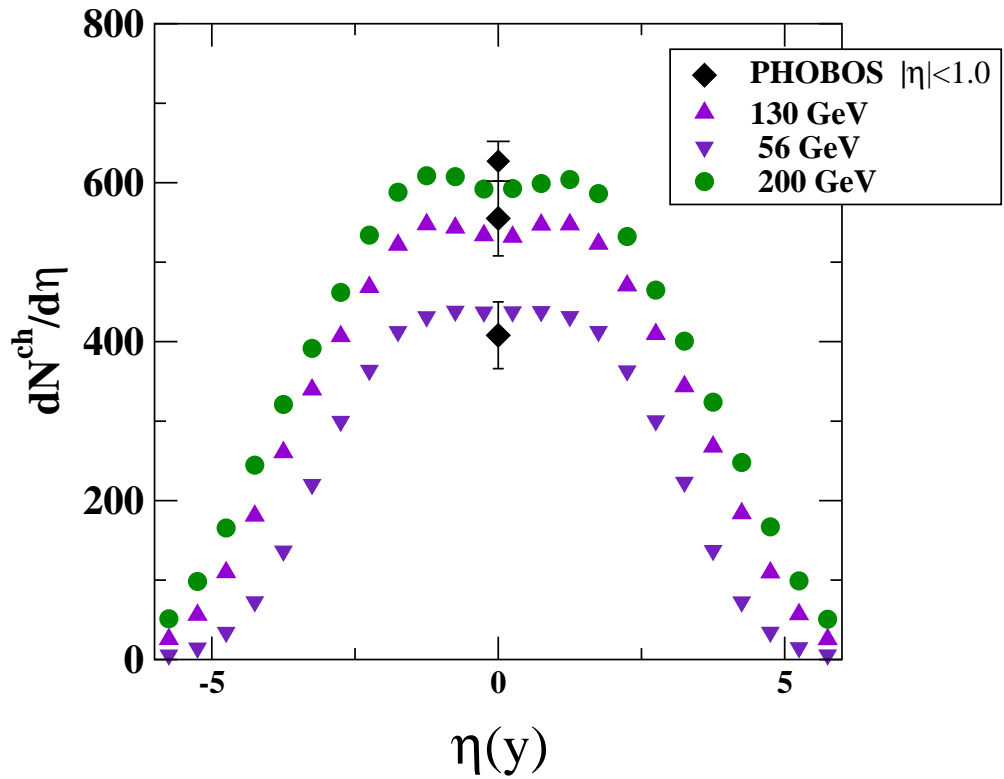}
\hfil}}
\caption[]{Previously    obtained    comparison    with    PHOBOS
$\sqrt{s}=56$ and  $130$ GeV $dN^{ch}/d\eta|_{\eta=0}$ indicating
what was  then a prediction for  the as yet  unmeasured $200$ GeV
run.  The latter  data now exists.  The small  relative rise from
$130$ to $200$  GeV in the $|\eta| < 1$ point  is a prediction of
the model.}
\label{fig:olddndeta}
\end{figure}

\begin{figure}
\vbox{\hbox to\hsize{\hfil
\epsfxsize=8.5truein\epsfysize=5.3truein\epsffile[35 58 750 750]
{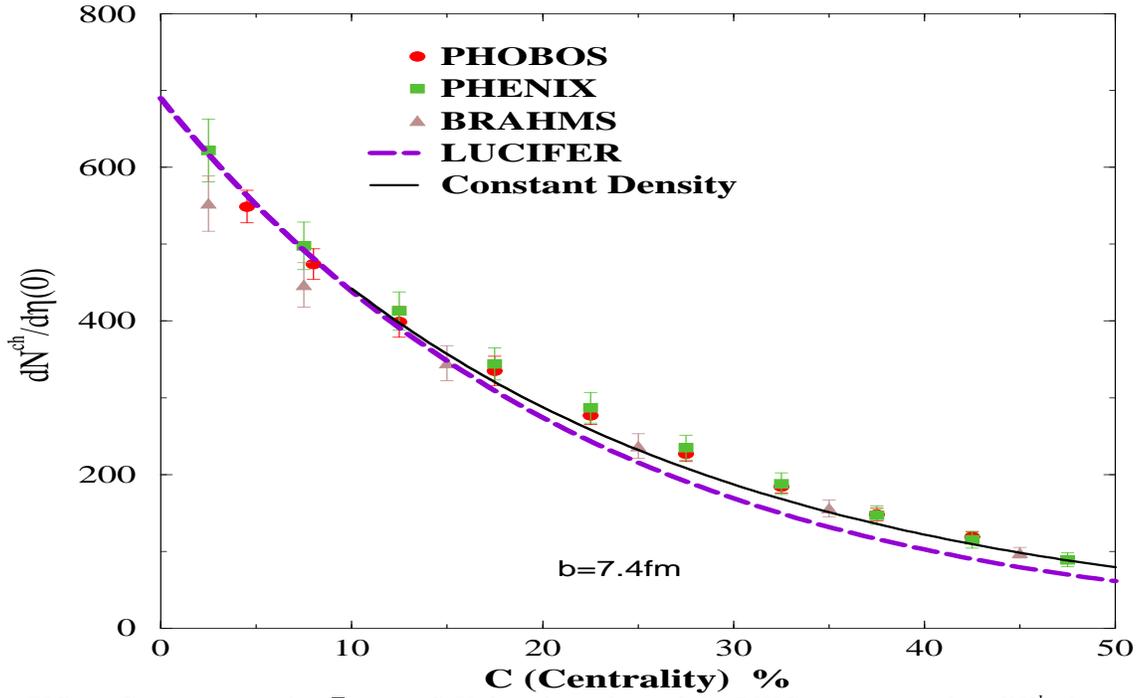}
\hfil}}
\caption[]{Comparison with $\sqrt{s}=130$  GeV data from three of
the RHIC  experiments for $(dN^{ch}/d\eta)(\eta)$ as  a function of
the  degree of  centrality.  The  latter is  defined  through the
percentage of  total cross-section measured,  experimentally, and
in the simulation by  a similar quantity determined geometrically
from the  impact parameter range $\Delta b$  spanned.  The dashed
line  is a  fit in  the form  $dN^{ch}/d\eta =  690  \exp \lbrace
-.0445C-.00008C^2\rbrace$.  The  upper solid curve  indicates the
result  obtained  if the  transverse  areal  density, defined  in
Eqn.~(3) and displayed in Fig.\ref{fig:sigma}, is held constant.}
\label{fig:dndetacentral}
\end{figure}

\begin{figure}
\vbox{\hbox  to\hsize{\hfil  
\epsfxsize=6.0truein\epsffile[0 0 600 600]{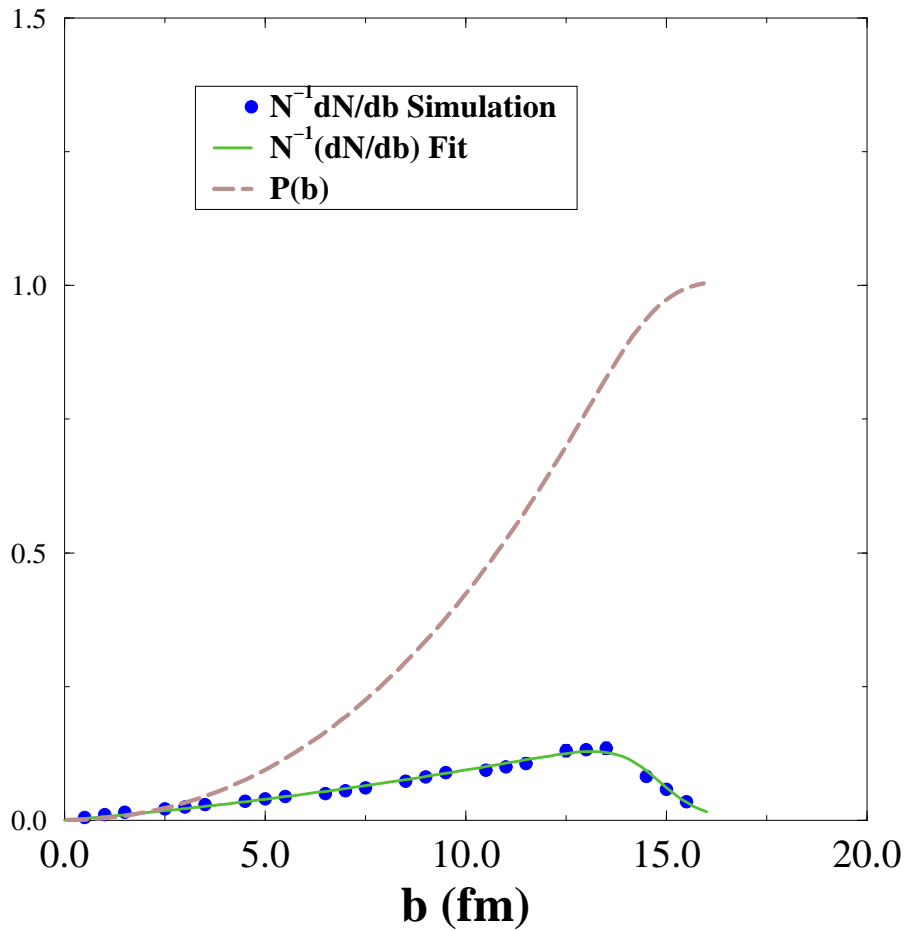} 
\hfil}}
\caption[]{Centrality vs Impact parameter in LUCIFER at $\sqrt{s}
= 130$  GeV.  The  lower curve  is a fit  to the  simulation data
points in the form $N^{-1}dN/db = \frac{1}{160} [b +.05 b^2]/[1 +
\exp \lbrace 1.8 (b-14.6)\rbrace] $ while the upper curve, giving
the  probability distribution  $P(b)$  of centrality  vs $b$,  is
obtained by integration.}
\label{fig:bvscent}
\end{figure}

Figure         \ref{fig:olddndeta}         reviews        earlier
$(dN/d\eta)^{ch}|_{\eta=0}$   results   for   the  most   central
collisions,  indicating the  predicted  $12$-$14$\% rise  between
$\sqrt{s}=130$   and   $200$   GeV   for   Au  +   Au   at   RHIC
\cite{lucifer2,phobos2}. A more complete picture of the variation
in  this pseudo-rapidity density  $(dN/d\eta)^{ch}|_{\eta=0}$ for
varying     centrality     cuts     is    shown     in     Figure
\ref{fig:dndetacentral}  with LUCIFER  compared  to the  existing
detector data  \cite{phenix1,star1,phobos2,brahms2} for the $130$
GeV runs.  The  centrality cuts on the simulation  are defined in
terms of the calculated $dN/db$,  which is integrated to give the
appropriate degree of centrality  between limits $b_1$ and $b_2$,
$\Delta b  = b_2 -  b_1$. Theoretical curves describing  this $b$
dependence  are  shown   in  Figure  \ref{fig:bvscent}.   A  more
expeditious definition of  the percentage centrality was employed
earlier\cite{lucifer2}. This convenient geometrical definition is
not  necessarily in  accord with  actual experimental  usage. The
agreement  between  simulation and  measurement  persists out  to
rather high impact parameter, possibly deteriorating somewhat for
$b  \ge 8$  fm.   For peripheral  collisions  the `constancy'  of
resonance density,  discussed in detail below,  is not maintained
in  our saturated  simulations.  Indeed  for the  most peripheral
collisions, with say only  a few nucleons colliding, there should
be no multiplicity suppression.

\begin{figure}
\vbox{\hbox to\hsize{\hfil
\epsfxsize=6.1truein\epsffile[25 28 584 750]{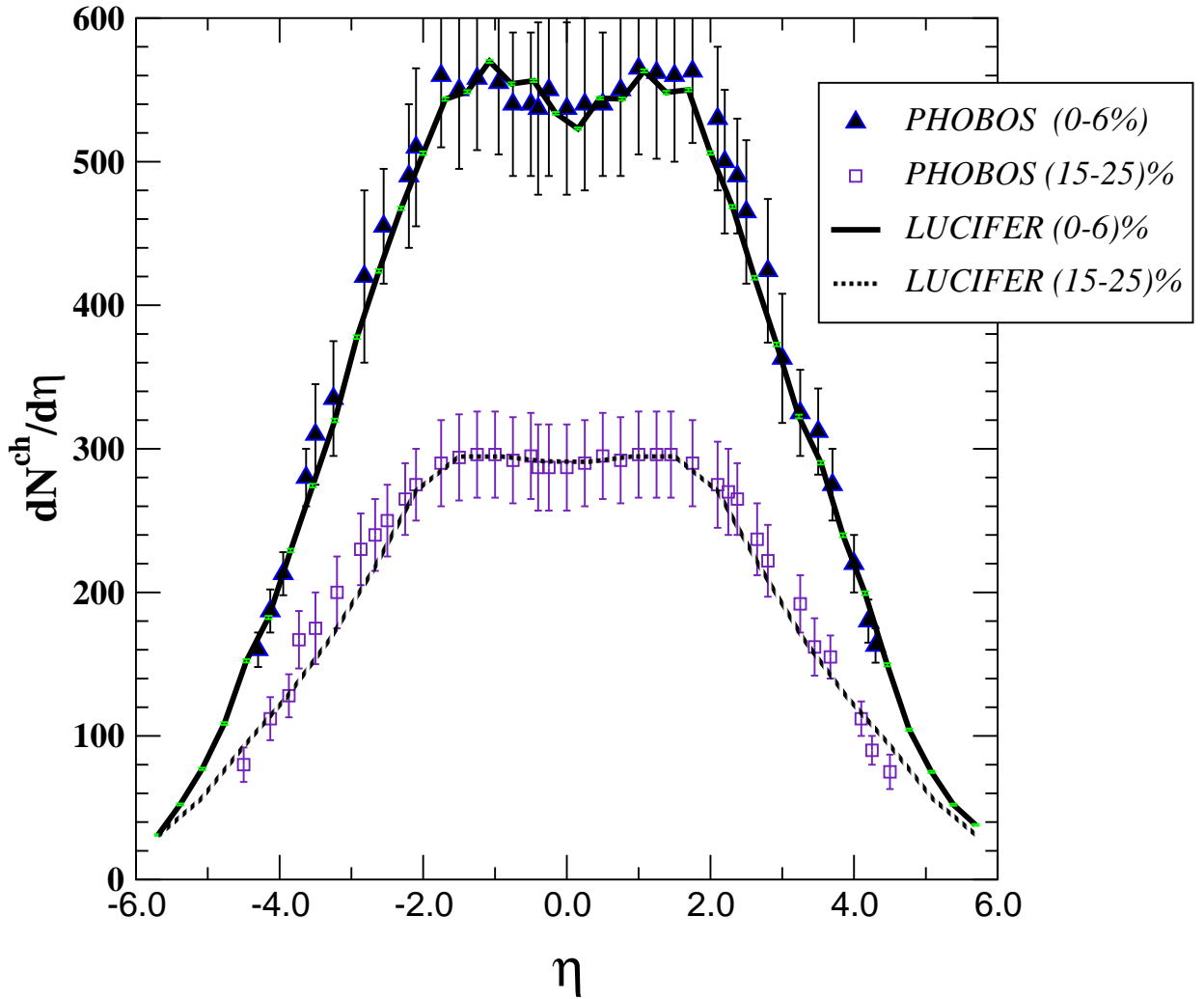}
\hfil}}

\caption[]{Full     pseudo-rapidity    distributions     for    a
representative  selection of  centrality cuts  at $\sqrt{s}=130$.
The calculation  for centrality of $20-25$\%  has been normalised
upward     by     $\sim     10$\%,    consistent     with     the
theoretical-experimental  centrality dependence  shown  in Figure
\ref{fig:dndetacentral}.}

\label{fig:pseudorapf}
\end{figure}

\begin{figure}
\vbox{\hbox to\hsize{\hfil
\epsfxsize=6.1truein\epsffile[0 0 584 750]{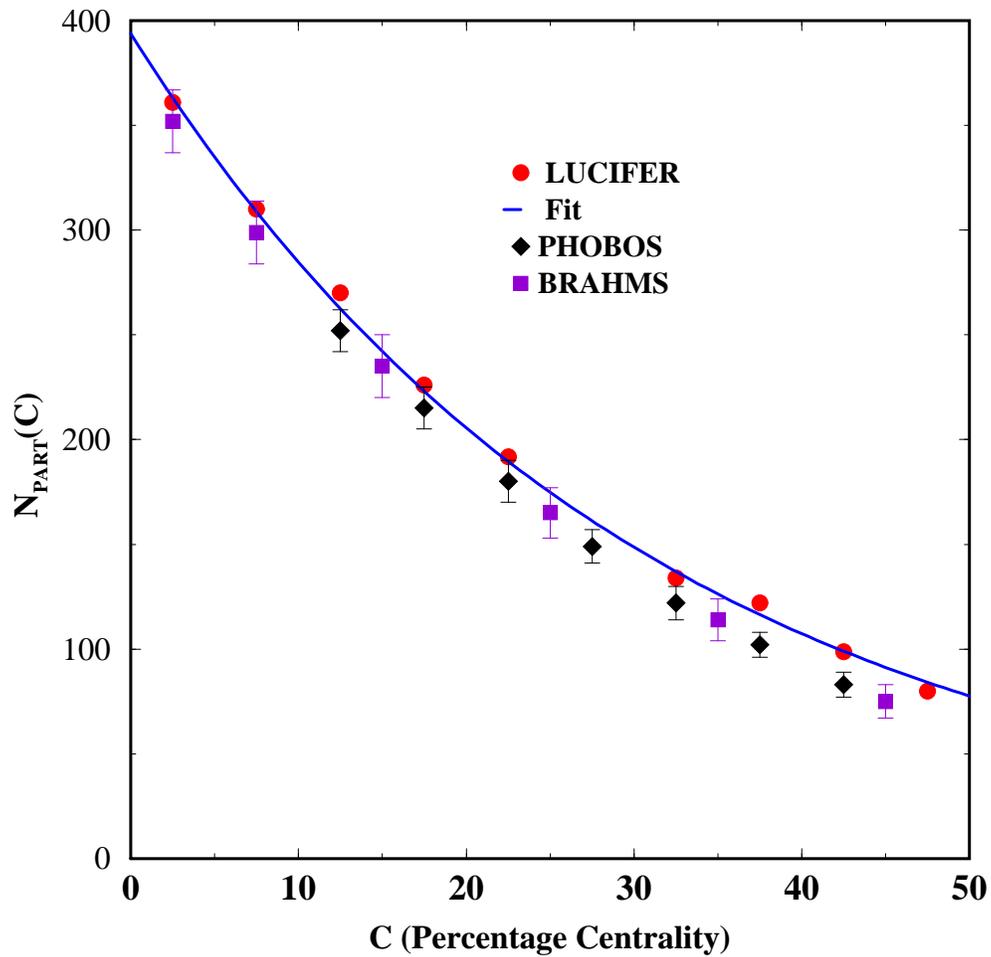}
\hfil}}
\caption[]{Participant  numbers from  LUCIFER against  those from
BRAHMS  and PHOBOS  as a  function of  centrality. A  fit  to the
simulation in the form $N^{part}=394 \exp \lbrace -.03355C\rbrace
$ is also displayed. }
\label{fig:partcomp}
\end{figure}

One should  also confront the  full pseudo-rapidity distributions
insofar  as  they  have  been  measured; this  we  do  in  Figure
\ref{fig:pseudorapf}  using  PHOBOS  data at  $\sqrt{s}=130$  GeV
\cite{phobos2}.  The higher  energy $\sqrt{s}=200$ yields similar
comparisons. Again the agreement in shape is seen to be adequate.
So the most inclusive meson data, taken at $\sqrt{s}=130$ GeV, is
well  described by the  present two  phase cascade.   Finally, to
demonstrate the  similarity with other presentations  of the data
we  include  a plot,  Figure  \ref{fig:partcomp}, of  participant
nucleon  number,  $N^{part}(inelastic)$,  in LUCIFER  versus  the
`number  of participants' which  was abstracted  from two  of the
experiments. In  the simulation we can easily  estimate the total
$N^{part}(inelastic)$  by   counting  the  number   of  spectator
nucleons  (they are  just  those initial  nucleons  which had  no
inelastic collisions in phase I or II of the cascade) in a sample
of events having some fixed  $b$ and subtracting that number from
the total  number of initial  nucleons. Of course  only inelastic
participants,  which  contribute  to  the multiplicity,  must  be
selected  for comparison  with experiment  where  the distinction
between  `inelastic'   participants  and  `elastic'  participants
becomes  moot,  so that  another  subtraction  of purely  elastic
participants  must  be  made  theoretically. For  the  degree  of
centrality  reached   in  the  experiments,   $\leq  50\%$,  this
additional  adjustment   is  small,   but  it  would   become  an
appreciable   correction,  tending  to   $\sim  20\%$,   in  more
peripheral events.

\section{Resonance Multiplicity Dependences on $b$ and $\sqrt{s}$} 

\subsection{Fire Cylinder and Transverse Areal Density}

\begin{figure}
\vbox{\hbox to\hsize{\hfil
\epsfxsize=6.1truein\epsffile[0 0 584 700]{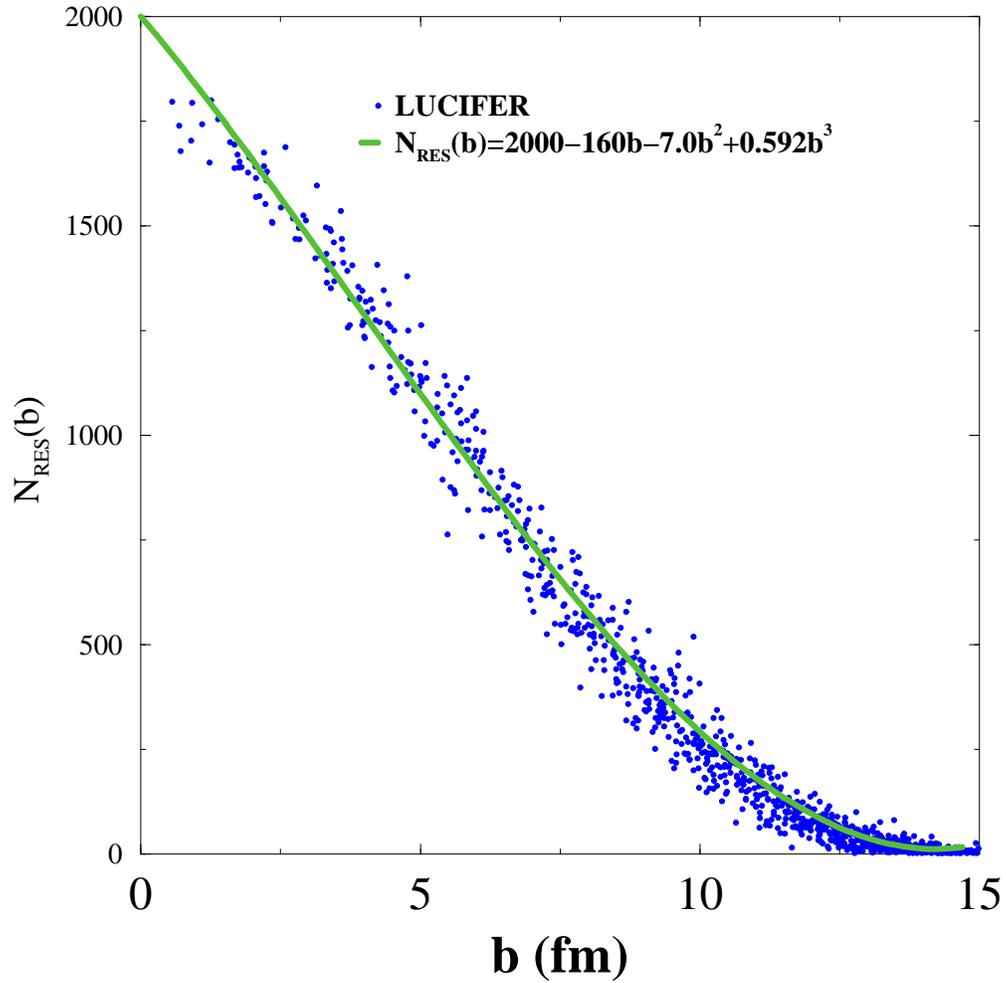}
\hfil}}
\caption[]{The   calculated  distribution   of   generic  bosonic
resonances with  impact parameter  at the end  of Phase  I. These
resonances  are generated  in Phase  I and  scatter and  decay in
Phase II.  The particle and  energy densities resulting  from this
distribution are discussed in the text.}
\label{fig:RESb}
\end{figure}

Most  of the  simulation results  can be  understood in  terms of
simply constructed  model quantities. At the  commencement of the
`soft'  cascade, phase  II, we  constrained the  system  to avoid
spatial  overlap of the  rescatterers. The  extent to  which this
succeeds  can be quantified  as a  function of  impact parameter.
Plotted  in  Figure \ref{fig:RESb}  is  the  distribution of  the
$\rho$-like  resonances as  a function  of $b$,  at the  start of
phase  II.   There is  of  course  a  dispersion in  the  numbers
$N_{RES}(b,s)$, reflecting fluctuations  from event to event, but
a good estimate  of the areal density of  resonances can still be
made. We use the term  `areal' density in referring to the number
of resonances in the area  of overlap of the two nuclei, obtained
when the  overlap volume from phase  I is projected  into a plane
transverse to the beam direction, for fixed impact parameter.

The overlap  volume or interaction region has,  in our simulation
of  a gold-gold  collision, for  early  times $ t > \tau_f$  very
nearly the form of a  generalised cylinder. The directrix of this
cylinder  is a  convex  curve  formed by  the  two circular  arcs
bounding the  projection of the  overlap volume of  the spherical
nuclei into  the transverse plane. The generator  of the cylinder
is  parallel to  the beam  direction and  along the  direction of
longitudinal expansion  of the interaction  region.  Perhaps this
volume could  be appropriately be termed a  `fire cylinder.'  The
area  in question then,  is the  transverse cross-section  of the
fire cylinder. Taking the geometry of the two colliding nuclei to
be sharply spherical,  with radius R, the area  of overlap $A(b)$
can be calculated:

\begin{equation}
A(b) =
2R^2\left[
cos^{-1}\left(\frac{b}{2R}\right)
- \left(\frac{b}{2R}\right)\sqrt{1-\left(\frac{b}{2R}\right)^2}
\right].
\end{equation}

\noindent For $b \leq R$, the areal density of resonances is then
found to be:

\begin{equation}
\sigma_{a}(b,s) = N_{RES}(b,s)/A(b).
\end{equation}

\begin{figure}
\vbox{\hbox to\hsize{\hfil
\epsfxsize=6.1truein\epsffile[0 0 584 700]{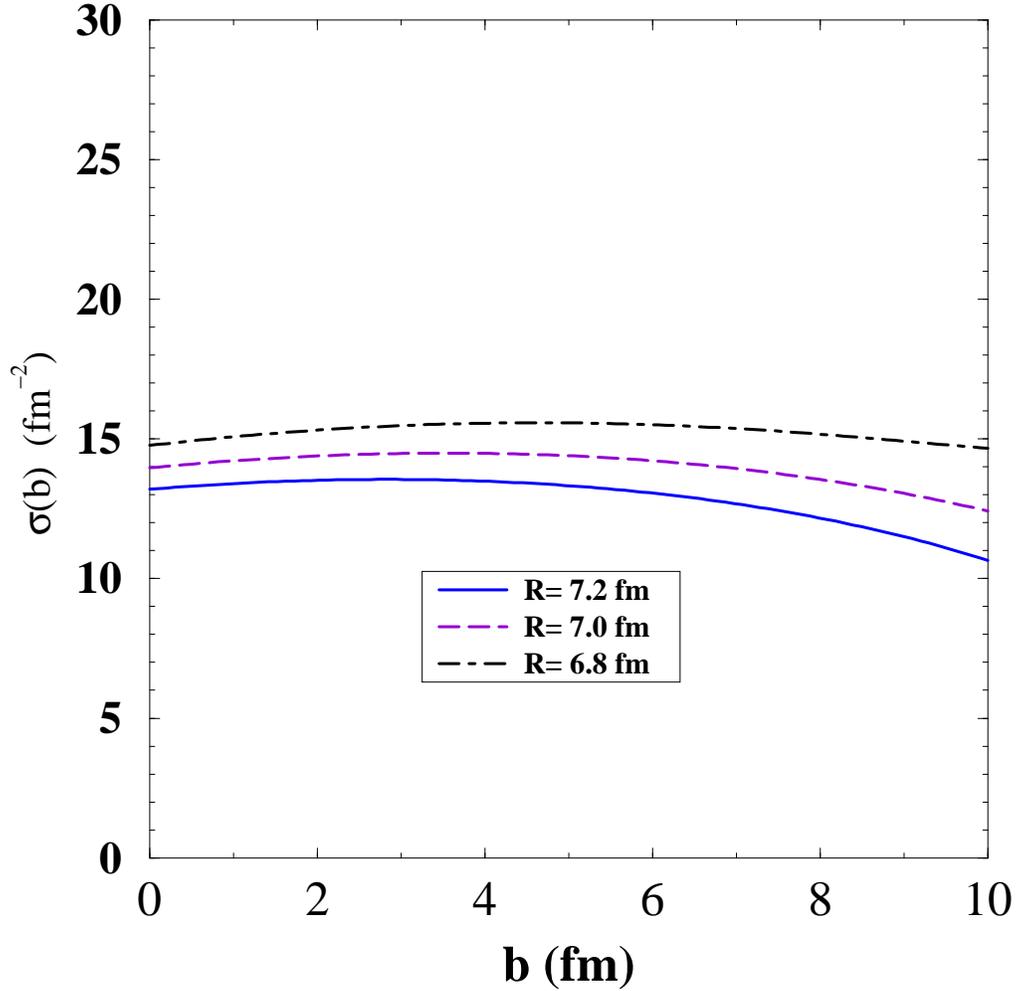}
\hfil}}
\caption[]{The  transverse  areal  density  $\sigma_{a}(b,s)$  of
generic  resonances as  function  of impact  parameter at  three,
radii $R$ for  colliding Au nuclei, assumed to  be sharp spheres.
The most  evident feature is  the slow variation of  this density
$\sigma_{a}$ with $b$.   For the largest $R$ there  is a residual
$\sim 12$\% drop in $\sigma_{a}$ from $b=5$ to $b=10$ fm.}
\label{fig:sigma}
\end{figure}

Sample curves  obtained for this density are  exhibited in Figure
\ref{fig:sigma}   for   varying   choices   of   nuclear   radius
$R=6.8,7.0,7.2$ fm,  the latter value being closer  to the actual
situation  for gold  nuclei, especially  when the  addition  of a
surface diffusivity converts the nuclear geometries into the more
realistic  forms  we  employ  in the  Monte-Carlo.   This  figure
demonstrates, however, the  near constancy of $\sigma_{a}(b)$ out
to  rather  large separations  of  the  colliding nuclei,  $b\sim
8$--$10$ fm. This  perhaps surprisingly accurate, but intuitively
satisfying, uniformity  of the  resonance population of  the fire
cylinder  is  a  representation   of  the  constraints  placed  on
production.   The result  of imposing  constancy strictly  out to
$b=10$     fm    is    included     in    the     above    Figure
\ref{fig:dndetacentral}.  Measurements seem to be telling us that
indeed  the  transverse  density  $\sigma_{a}(b,s)$  varies  very
little.

A reasonable  approach for still more peripheral  events would be
to smoothly  cut off the  degree of suppression of  the resonance
multiplicity, until for the most glancing collisions there was no
multiplicity suppression at all, at which point one expects after
all that only unadulterated two body reactions occur. The results
for    the    latter   eventuality    are    shown   in    Figure
\ref{fig:ua5dndeta}, as explained  previously a prediction of the
elementary two-body model used  here and extracted, in fact, from
p+D reactions.

\subsection{Fire Cylinder Volume and Energy Dependency}

\begin{figure}
\vbox{\hbox to\hsize{\hfil
\epsfxsize=6.1truein\epsffile[27 58 584 700]{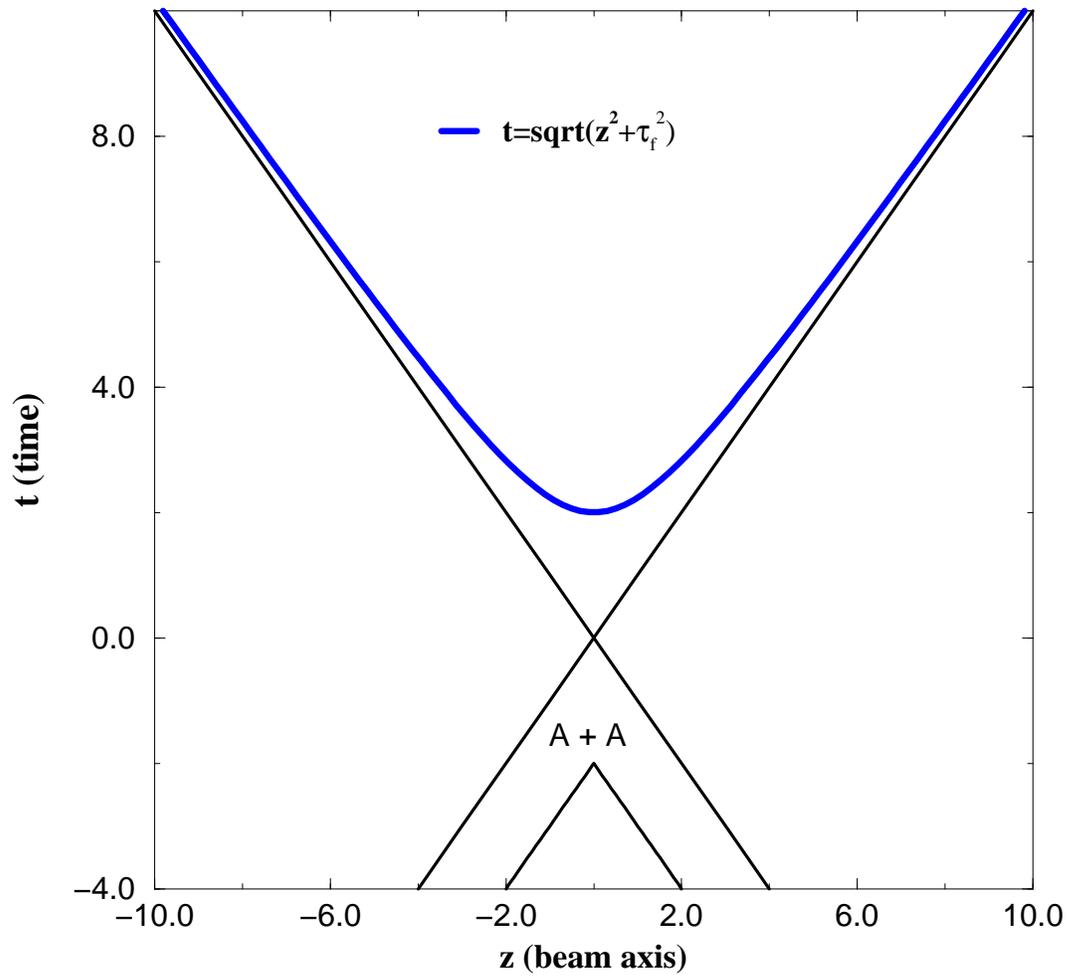}
\hfil}}
\caption[]{A graphical  space-time representation of  the ion-ion
collision. The second phase of the cascade begins along the solid
upper curve  shown in the figure.   This curve is  defined by the
formation  time  $\tau_f$,  before  which the  generic  bosonic
resonances are not in  play.  Only longitudinal motion, along the
beam axis, is represented. }
\label{fig:lightcone}
\end{figure}

It  is  more difficult  to  precisely  define  a volume  for  the
interaction region or  fire cylinder at the onset  of phase II of
the simulation, the soft cascade. From figure \ref{fig:lightcone}
it  is clear  that  after time  $t  = \tau_f$  in  the center  of
momentum (lab) frame, up to  the time of termination of phase II,
new  resonances are  continually appearing,  at times  related to
their velocities $\beta_i$ via $t_i= \gamma_i \tau_f$, $\gamma_i$
being the Lorentz factor peculiar to the $i$'th resonance. In the
final  analysis,  a  choice  $\tau_f  =  1.2-1.6$  fm/c  for  the
formation  time  was  made,  with globally  similar  results.  In
general, it is clear there is some interplay between $\tau_f$ and
the overall  normalisation of  resonance number. A  more extended
delay time implies  a more expanded system and  less cascading in
phase II.   Too short a formation  time could in fact  lead to an
inability  for the  simulation  to keep  within the  experimental
limits for  total meson numbers.  No such problem arises  for the
range of  $\tau_f$ selected above,  and these total numbers  as a
function of centality are well reproduced.

\begin{figure}
\vbox{\hbox to\hsize{\hfil
\epsfxsize=6.1truein\epsffile[27 58 584 700]{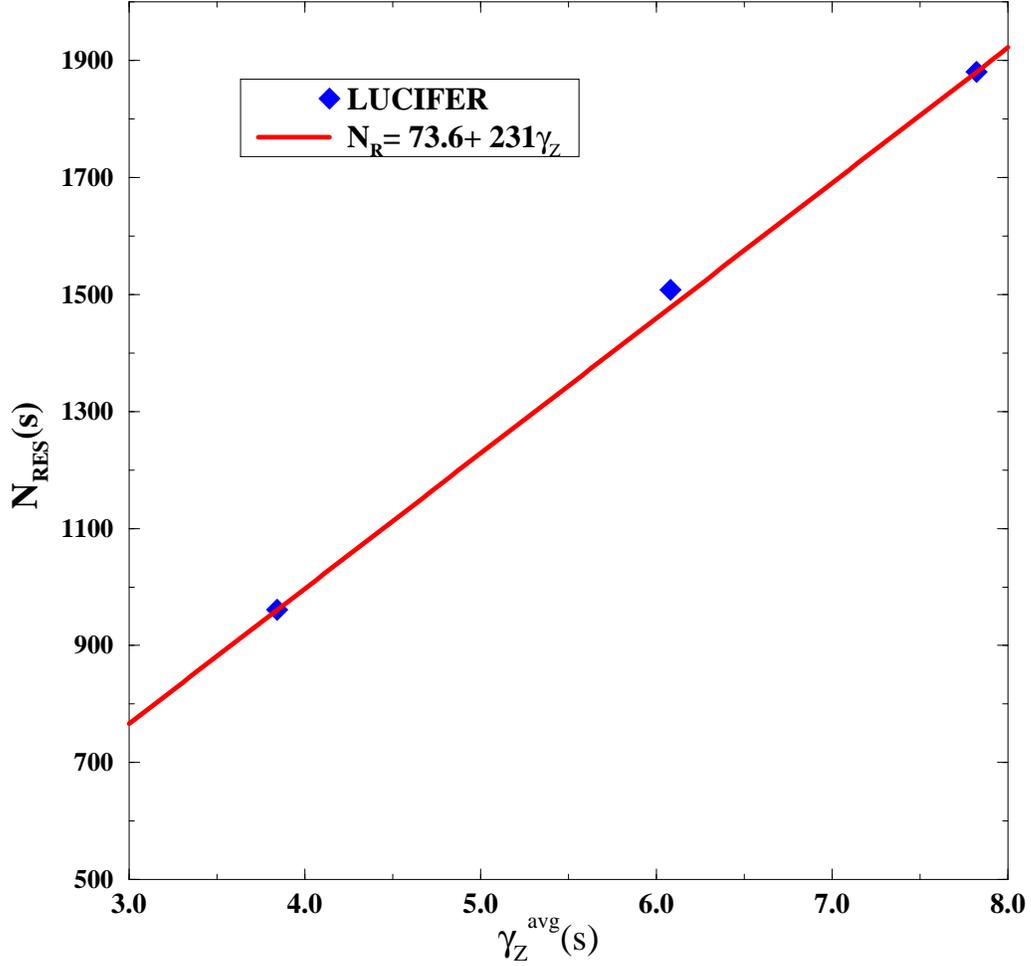}
\hfil}}
\caption[]{Dependence     of    calculated     $N_{RES}(s)$    on
$\overline{\gamma_z}(s)$ for  $\sqrt{s} = 56, 130,  200$ GeV. The
close to linear relationship  between these variables defines the
variation in  final meson  multiplicity with energy.   The growth
with $s$ derives  from the input KNO scaling and  leads to a near
constant, or saturated, spatial density.}
\label{fig:gammaz}
\end{figure}

For these  last issues  of spatial densities  we choose  the most
central  collisions with  $b \leq  4$, although  little  would be
changed   by   using    a   fixed   impact   parameter.    Figure
\ref{fig:gammaz} demonstrates  the strong interdependence between
the   number   of  resonances   $N_{RES}(s)$   and  the   average
longitudinal  Lorentz  factor  $\overline  {\gamma_z}  (s)$  (the
averaging here  is over all  resonances).  $\overline {\gamma_z}$
is directly related to the average length of the fire cylinder in
the longitudinal direction:

\begin{equation}
L_z (s)= 2 \tau_f \overline{\gamma_z}.
\end{equation}
\noindent  The energy  dependence  of $N_{RES}(s)$  is then  very
simply and naturally described in Figure \ref{fig:gammaz}, and is
another dynamical  statement of  the model.  The  cylinder length
grows  linearly with  $\overline{\gamma_z}$  for increasing  $s$,
once again a useful and easily understood, result.

\noindent It is then an easy  matter to compute the volume of the
fire cylinder

\begin{equation}
V(b) = A(b) L_z (s),
\end{equation}

\noindent  and  thus  to  obtain  the  three  dimensional  number
density:

\begin{equation}
\rho(b,s)= \sigma_{a}(s)/{L_z(s)}.
\end{equation}

For this selection of  very central collisions $\rho(s)$ is found
to  be $\sim 0.70-1.10$  particles per  fm$^3$, depending  on the
specific  choices  for  nuclear  radius and  $\tau_f$  in  Figure
\ref{fig:sigma}. This leads to an average resonance size of $\sim
0.70$  fm. Of  course, in  the  simulation we  treat the  generic
scatterers  as  points  and  their  extracted size  is  simply  a
statement of the volume each  resonance occupies, more or less at
the time it is created.   It is, however, significant that such a
size is  eminently reasonable for a hadronic  resonance, and that
the  associated number  density is  consequently  relatively low.
These averages obtain of course after the formation time $\tau_f$
in the rest  frame of produced mesons, but  considerably later in
say the c. of m.  Nevertheless, to use an earlier, `average' time
would also involve less produced particles, and the fire cylinder
densities we introduce would be little influenced.

\section{Transverse Number and Energy Densities}

One   can  also   estimate  the   transverse   energy  densities,
$\rho_{\epsilon}(t)$,  associated with  the above  number density
$\rho(b,s)$.   Considering, for  now, only  the  dominant bosonic
contribution,  one arrives at  a total  $\sim \sqrt{(1)^2+(1)^2}$
GeV {\it i.~e.~}  $\sim 1.4$ GeV/ fm$^3$. This  since the average
bosonic mass is  somewhat over $1$ GeV and  its average $p_\perp$
in phase  I is somewhat under  $1$ GeV/c.  To this  must be added
the  energy density  from baryons  present at  the  initiation of
phase II, an increase within  the fire cylinder imagined above of
some   $20$\%;  thus   overall   $\rho_{\epsilon}(t)  \sim   1.7$
GeV/$fm^3$.  The  density $\rho_\epsilon (t)$ is  an average over
many events  constructed at  the start of  phase II, and  is well
below the  average Bjorken value quoted by  the RHIC experimental
collaborations.  Almost  independent of collider  energy, a major
proportion, some  $40\%$ of  the final mesons,  and hence  also a
large  fraction of  the transverse  energy, is  produced  via the
collisions  taking  place  in  phase II  \cite{lucifer2}  of  the
simulation. The variation in this  fraction is only a few percent
from $56$ to $200$ GeV/A.  Combining the contributions from I and
II would yield $\sim 2.8$  GeV $fm^3$, still short of the Bjorken
estimate.  Part  of this  discrepancy in total  transverse energy
densities is  explained by  our use of  a longer  formation time.
However, since  the final  phase of the  cascade occurs  when the
system  has  surely expanded  considerably  spatially, the  extra
transverse energy generated in II contributes little to the early
density, often used as a  trip wire in theoretical models for the
initiation of `plasma' creation.

\begin{figure}
\vbox{\hbox to\hsize{\hfil
\epsfxsize=6.1truein\epsffile[0 0 584 750]{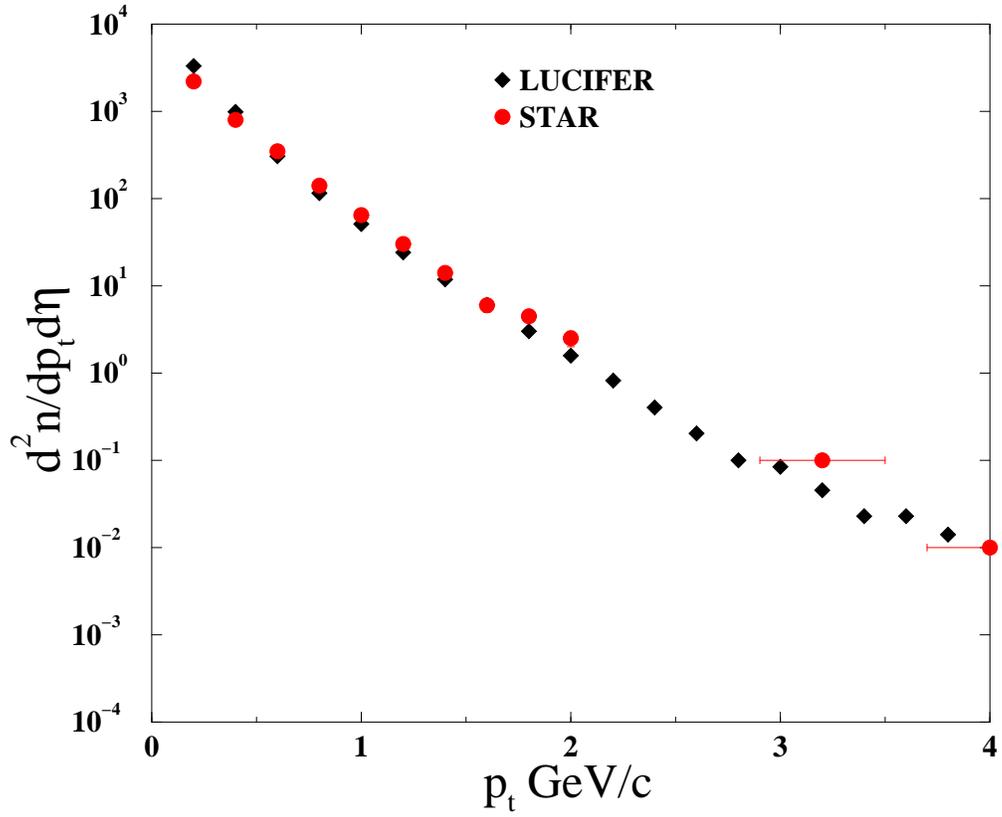}
\hfil}}
\caption[]{Transverse   momentum    distribution.    LUCIFER   vs
STAR. The  experimental data  includes all negative  hadrons; for
the   theory   anti-protons   are   omitted,  but   if   included
\cite{phenix2} they would make the agreement even closer for $p_t
> 2$ GeV/c where $\bar p$ becomes appreciable.}
\label{fig:highpt}
\end{figure}

The  question  of  detailed double  differential  cross-sections,
$d^2N/d\eta  p_\perp dp_\perp$, especially  those at  the highest
transverse  momenta  measured,  will  be examined  thoroughly  in
ensuing  work. For now  we present  in Figure  \ref{fig:highpt} a
preliminary  comparison   between  STAR  data   \cite{star2}  and
LUCIFER, pertaining  to total negative  hadron ($h^-$) production
below $p_\perp = 4 GeV/c$.  The higher ranges of this comparison,
say $p_\perp  \ge 2$ GeV, are  possibly beyond the  limits of our
current  modeling which  largely  omitted hard  processes in  the
underlying two body scattering.  Large $p_\perp$ events certainly
require  more study,  but  the present  results are  nevertheless
interesting.  The highest $p_\perp$ particles are produced in the
simulation by multiple scattering  and are thus subject to normal
energy  losses   expected  in  such   scattering.   Despite  this
reservation, the  rapid drop in  $d^2N / d\eta  p_\perp dp_\perp$
assures one that the softer  part of the spectrum dominates total
transverse  energy production. Thus  the total  transverse energy
$E_t$   generated  in  the   cascade  is   close  to   that  seen
experimentally.

Of course  we cannot really  say what transpires in  the earliest
stages  of the ion  collision. We  have made  certain assumptions
about hadron  tenacity and saturation  and have then  obtained at
the close of the second stage of the cascade conditions very much
like  those  observed   for  the  global,  inclusive,  production
processes. One  problem, mentioned in  the Introduction, concerns
the  earliest phase  of the  collision, perhaps  even  before the
production  of  bosonic   resonances,  and  involves  the  purely
baryonic  densities. If one  assumes each  nucleus is  never less
than $1$  fm in transverse  extent, then for $b=4$  fm collisions
the  baryon density  does not  rise  much above  one per  fm$^3$.

Further, the successful use of generic resonances to describe the
cascading  might  simply  reflect  a correct  fixing  of  average
cross-sections and cross-section `flow' during an ion-ion event.

\section{Comments and Conclusions.}

The two phases of the  LUCIFER simulation have been merged into a
single, idealized,  dynamical model. Such  a model can be  only a
part  of  the complete  story,  of  course;  partonic degrees  of
freedom do  exist and  for sufficiently high  particle transverse
momentum   presumably  will   dominate  the   inclusive  spectra.
Nevertheless,  in our approach  centrality and  energy dependence
comparisons between data and simulation are understood as arising
from a common physical  origin: the uniform and unvarying density
of resonances in  the fire cylinder.  The constancy  of areal and
volume densities of bosonic  resonances and their close to linear
dependence  in  energy  with  $\overline{\gamma_z}$  are  notable
results of the simulation.

The  dynamics  of  the   collision  is  split  into  two  phases,
sequential in time: one could have created the initial conditions
for  phase   II  in  some  alternate  fashion,   for  example  by
attributing the initial meson  distributions to a string model or
to     de-confinement     and     hadronisation    of     partons
\cite{werner,gluesaturation}. Our sole and central point here is,
then, that signals  other than those so far  observed should show
up  to distinguish  the present  cascade from  such alternatives,
especially from nearly complete parton de-confinement. The string
approach may  in fact  lead to the  same results in  simulating a
massive  ion-ion  collision.   We  reiterate  that  the  two-body
modeling and cascading in phase I leaves strong footprints in the
final results, especially in  the surprisingly gentle increase in
predicted, and measured,  $(dN/d\eta)^{ch}|_{\eta = 0}$ seen from
$\sqrt{s}=130$ to $200$ GeV.

No one  can doubt that there  is a saturation  effect observed in
meson  production at  RHIC.  To  produce this  effect,  we reduce
cascading in two  fashions: (1) through the very  presence of the
boson resonances  which eventually decay, over both  phases I and
II, into many  more $\pi$ or $K$ mesons  ($\sim 4$--$5$), and (2)
through the near constant  resonance density in the fire cylinder
achieved in  central collisions.  This  latter can be  imposed in
principle via, and is equivalent to, placing a no spatial overlap
constraint on the produced  resonances prior to the soft cascade.
The data,  Figure \ref{fig:dndetacentral}, suggest  a more strict
uniformity than produced in the cascade.

In any case, the relatively small observed total multiplicities,
which present a problem for all models, do not strongly support a
large increase in the number of degrees of freedom in the early
phase of Au + Au collisions at RHIC.

One should add that no attempt was made here to divide the output
of  the   cascade  into   separate  dependences  on   numbers  of
participants or on the  numbers of binary collisions. The cascade
does this  dynamically and  seamlessly. One might  note, however,
that  the  participant  number  directly determines  the  overall
incoming energy  put into  an ion collision  in phase I  and thus
also  fixes  the number  of  resonances  at reinitialisation  and
commencement  of phase II.   The division  into two  phases does,
however, distinguish the present approach from some alternatives;
principally in  reducing the  generation of transverse  energy at
very early times.

We will  return in  later work to  more specific  questions, {\it
e.~g.}~of  flavour   generation  and  high   transverse  momentum
spectra, but we mention here  that close to the correct ratios of
strange  meson production at  mid rapidity,  are obtained  in the
calculations.   These  contribute  some  $10$\% of  the  particle
number there.

This work  addresses only  gross, inclusive features  of existing
RHIC   data.   Accordingly,  the   results   only  suggest   that
non-standard medium effects must be looked for in more exclusive,
rare events,  whose analysis we  all await. Places to  look might
include   still  higher   transverse   momenta  spectra,   direct
$\gamma$-rays, $\mu^{+}\mu^{-}$  and $e^+e^-$ pairs,  or possibly
just a sampling of  events containing unusually high multiplicity
fluctuations.

\section{Acknowledgements}
This  manuscript  has  been  authored  under  the  US  DOE  grant
NO. DE-AC02-98CH10886. One of  the authors (SHK) is also grateful
to the  Alexander von Humboldt Foundation, Bonn,  Germany and the
Max-Planck Institue for Nuclear Physics, Heidelberg for continued
support  and  hospitality.  Useful  discussion with  the  BRAHMS,
PHENIX,   PHOBOS   and   STAR   collaborations   are   gratefully
acknowledged.

\clearpage

\clearpage

\clearpage

\clearpage

\clearpage

\clearpage

\clearpage

\clearpage

\clearpage

\clearpage

\clearpage

\end{document}